\documentclass[fleqn,usenatbib]{mnras}

\usepackage{newtxtext,newtxmath}

\usepackage[T1]{fontenc}
\usepackage{ae,aecompl}

\usepackage{graphicx}	
\usepackage{amsmath}	
\usepackage{amssymb}	
\usepackage[normalem]{ulem} 
\usepackage{tikz}
\usepackage{soul}
\usepackage{ wasysym } 
\newcommand*{\swap}[2]{#2#1} 

\title[Dust evolution in the disc of HD 50138]{Dust evolution in the circumstellar disc of the unclassified B[e] star HD 50138}

\author[J. Varga et al.]{
J. Varga,$^{1, 2}$\thanks{E-mail: varga@strw.leidenuniv.nl}
T. Gerj\'ak,$^{2, 3}$
P. \'Abrah\'am,$^{2}$
L. Chen,$^{2}$
K. Gab\'anyi,$^{4, 2}$
and
\'A. K\'osp\'al$^{2, 5}$ 
\\
$^{1}$Leiden Observatory, Leiden University, PO Box 9513, NL2300, RA Leiden, the Netherlands \\
$^{2}$Konkoly Observatory, Research Centre for Astronomy and Earth Sciences, Hungarian Academy of Sciences, \\ Konkoly Thege Mikl\'os \'ut 15-17., H-1121 Budapest, Hungary\\
$^{3}$Department of Astronomy, E\"otv\"os Lor\'and University, P\'azm\'any P\'eter s\'et\'any. 1/A, Budapest, Hungary\\
$^{4}$MTA-ELTE Extragalactic Astrophysics Research Group, P\'azm\'any P\'eter s\'et\'any 1/A, H-1117 Budapest, Hungary \\
$^{5}$Max Planck Institute for Astronomy, K\"onigstuhl 17, D-69117 Heidelberg, Germany\\
}

\date{Accepted XXX. Received YYY; in original form ZZZ}

\pubyear{2018}

\begin{document}
	\label{firstpage}
	\pagerange{\pageref{firstpage}--\pageref{lastpage}}
	\maketitle
	
	\begin{abstract} 
		We studied the disc of the unclassified B[e] star HD 50138, in order to explore its structure, and to find indications for the evolutionary status of this system, whether it is a young Herbig Be or a post-main-sequence star. {Using high spatial resolution interferometric measurements from MIDI instrument (N-band) on the Very Large Telescope Interferometer}, we analysed the disc size, the time-variability of the disc's thermal emission, and the spectral shape of the $10~\mu$m silicate feature. By fitting simple disc models, we determined the inclination and the mid-infrared size of the disc, confirming earlier results based on a lower number of observations. We searched for mid-infrared temporal variability of different regions of the disc, and concluded that it{s morphology} is not experiencing significant changes over {the observed epochs}. We characterized the mid-infrared silicate feature by determining the feature amplitude and the $11.3/9.8~\mu$m flux ratio. The latter parameter is a good indicator of the grain size. The shape of the feature suggests the presence of crystalline silicate grains in the disc. The interferometric data revealed a strong radial trend in the mineralogy: while the disc's innermost region seems to be dominated by forsterite grains, at intermediate radii both forsterite and enstatite may be present. The outer disc may  predominantly contain amorphous silicate particles. A comparison of the observed spectral shape with that of a sample of intermediate-mass stars (supergiants, Herbig Ae/Be stars, unclassified B[e] stars) implied that the evolutionary state of HD 50138 cannot be unambiguously decided from mid-IR spectroscopy.
		
	\end{abstract}
	
	
	
	
	\section{Introduction}
	\label{sec:intro} 
	
	B[e] stars are a class of stars with B spectral type, showing forbidden optical emission lines and infrared (IR) excess, which is emitted by circumstellar dust \citep{AllenandSwings1976}.
	The B[e] phenomenon is still not very well understood, despite a long history of observations. B[e] stars constitute a heterogeneous group regarding stellar evolution, divided into five main subclasses: supergiants, pre-main sequence (Herbig B[e]) stars, compact planetary nebulae, symbiotic stars, and unclassified B[e] stars \citep{Lamers1998}. Currently half of the B[e] stars are unclassified \citep{Fernandez2011}, however, high spatial resolution observations can provide an efficient classification tool for them \citep{Miroshnichenko2007}. A notable feature of unclassified B[e] stars is their {high} mass loss rates, {compared to normal main-sequence stars,} which can not be explained by the wind-theory for main-sequence B-type stars \citep{Fuente2015}. {Originally unclassified B[e] stars were thought to be isolated objects, but recently \citet{Fuente2015} reported that these objects can be also found in clusters. They estimated the age of the stars in the range from 3.5 to 6.5 Myr, which is out of the range for both pre- and post-main sequence origin, as the former phase lasts for $\sim 0.1$ Myr, and the main-sequence lifetime is $15-20$~Myr for a $12\ M_\odot$ star.}
	
	\citet{Miroshnichenko2007} revised the classification of B[e] stars, and proposed a new name, FS CMa stars, for the unclassified B[e] objects. He argued that these stars are binary systems undergoing rapid mass exchange, and that they are still on or close to the main sequence. The known fraction of binaries among unclassified B[e] stars is about $30$ per cent. Binarity could also explain the high observed mass loss rates.
	
	The forbidden lines in the spectrum of B[e] stars are originating from low density, extended circumstellar material. Other common properties in their optical spectrum are the strong Balmer emission lines and permitted emission lines of low ionization metals, emitted above the stellar photosphere by a $\sim 10^4$~K hot ionized gas. 
	The temperature of the dust around the star is about 500 to 1000K \citep{Lamers1998}{, but the presence of hot dust ($T \sim 1500$~K) has been also considered to model the K band continuum emission \citep{Ellerbroek2015}.} 
	There is substantial evidence that the dusty material around B[e] stars is geometrically distributed in a circumstellar disc. Herbig B[e] stars definitely have protoplanetary discs, formed by the collapsing protostellar cloud core. Other types of B[e] stars are also candidates for hosting circumstellar discs, which are formed from the ejected material from the star \citep{Oudmaijer2012}. {Moreover, new models reproducing forbidden emission lines and observations indicate that B[e] supergiants can also host circumstellar rings \citep{Kraus2016a,Kraus2016b}.}
	
	HD 50138 (aka. MWC 158) is a B6-7III-V[e]
	star \citep{Fernandes2009}, with a mass of $6-7\ M_\odot$ \citep{Fernandes2009,Ellerbroek2015}, at a distance of $377 \pm 9$~pc \citep{Gaia2018,Bailer-Jones2018}. {We adopted this distance value for the further analyses. HD 50138} has been classified as a classical Be star or as a Herbig Ae/Be star by several authors \citep{Fernandez2011}. \citet{Jaschek1993} stated that it is in a transitional state between a classical Be and a B[e] star. {Later} \citet{Lamers1998} labelled it as an unclassified B[e] star. Due to the difficulty of the classification, its evolutionary phase is not known. HD 50138 exhibits strong spectral variability, with time-scales lasting from days to months \citep{Pogodin1997}. \cite{Fernandes2009} analysed the spectroscopic variability of HD 50138 using high-spectral-resolution optical spectra from 1999 and 2007. They suggested that a new shell phase could have taken place before 2007. They determined the stellar parameters and got $\log(L_\star/L_\odot) = 3.06 \pm 0.27$ for the stellar luminosity {($2.81 \pm 0.27$, if  recalculated with the new Gaia DR2 distance)}. 
	\citet{Jerabkova2016} monitored this star over twenty years using optical spectroscopy. They confirmed the already reported quasi-periodic nature of the spectral variations (with a combination of two periods of $\sim$9 and $\sim$14 years). {They found only modest signatures for regular short term variability, but a 30 day period can be detected in the radial velocity measurements of \citet{Merrill1931}. Such a period can be consistent with the orbital period of a binary system with mass transfer between the components.} 
	
	The binarity of HD 50138 has been hypothesized by several authors. \citet{Baines2006} used spectro-astrometry to detect binary companions in a sample of intermediate-mass stars. They stated that the observed spectral signatures of HD 50138 are consistent with binarity. The separation of the companion is estimated to be in the range from $0.5\arcsec$ to $3\arcsec$. Near-IR interferometric observations also indicate asymmetries in the distribution of the circumstellar material, which can also be explained with the presence of a companion  \citep{Ellerbroek2015,Lazareff2017,Koutoulaki2018}.
	
	More recently, IR interferometry proved to be a powerful tool to reveal the structure of the circumstellar environment of B[e] stars \citep{Fernandes2010}. The existence of the circumstellar disc of HD 50138 has been directly confirmed by high resolution near- and mid-IR interferometric observations \citep{Fernandez2011}. If the star is a post-main-sequence object, then the disc is likely {formed by decretion or as a result of binary interaction} \citep{Ellerbroek2015}.  
	\cite{Fernandez2011} specified the geometrical parameters of the disc using near- and mid-IR interferometric observations from the VLTI/AMBER, VLTI/MIDI and the Keck segment-tilting experiment. They got $i=56\pm4^{\circ}$ for the inclination and $\theta=71\pm7^{\circ}$ for the position angle of the disc. \citet{Ellerbroek2015} observed the Br$\gamma$ line with VLTI/AMBER. They were able to resolve the disc, and found that the line emitting region is smaller {($< 2.3$~au, adopting the new Gaia DR2 distance)}, than the continuum emitting region. This is expected, as the line emission comes from the inner few au region, which is too hot for dust to exist. 
	
	
	\citet{Kluska2016} monitored HD 50138 with VLTI/PIONIER in H band over three epochs. They found that the morphology of the circumstellar environment is asymmetric, and varies over weekly or daily time scales. They explained this variability with a model which includes a disc and a bright spot which moves around the central star. Fitting the data with a circular trajectory they got $5.2$~mas {($=1.96$~au, adopting the new Gaia DR2 distance}) for the {radius}, and $P = 88$~days for the orbital period. The orientation and inclination of the fitted orbit is in agreement with the results of \citet{Fernandez2011}. The fit is not good enough to confirm the existence of a companion. They found it likely that the bright spot is actually a disc asymmetry, which could be caused by spiral arms or other disc features. 
	
	\citet{Lazareff2017} observed HD 50138 as a member of a large sample of Herbig Ae/Be discs in the H band using VLTI/PIONIER. They fitted models to the data to determine the geometry of the inner disc. They found $68-72^\circ$ for the position angle and $56-58^\circ$ for the inclination, which is in excellent agreement with the results of \citet{Fernandez2011} from mid-IR data. They also noted that disc asymmetry is indicated by the closure phases.
	Most recently, \citet{Koutoulaki2018} presented an interferometric study with VLTI/AMBER, revealing that the circumstellar environment is complex. They estimated a size of {$0.7-1.1$~au} for the continuum emitting region, and {$\sim$}0.4~au for the \ion{H}{i} emitting region\footnote{{Here we also recalculated the sizes using the recent Gaia DR2 distance measurement.}}. They also found, in agreement with the results of \citet{Ellerbroek2015}, that the Br$\gamma$ line-emitting gas is located in a Keplerian rotating disc. {These} authors noted that the observations indicate asymmetric disc structure, or maybe the presence of a companion. Although these IR interferometric studies revealed much of the rich structure of the circumstellar environment of HD 50138, they were unable to draw conclusions of the evolutionary state. Most authors, however, slightly favour the post-main-sequence scenario.
	
	\cite{Lee2016} studied eight isolated B[e] stars, including HD 50138. They argued that HD 50138 is a post-main-sequence star on the basis, that it is not associated with any known star forming region. They observed HD 50138 with the Submillimeter Array (SMA) at $1.3$~mm, and it was unresolved with an angular resolution of $\sim$$1\arcsec$ ($\sim$400~au). The steep fall-off of the spectral energy distribution at sub-millimetre wavelengths indicates a lack of cold grains, thus more compact distribution of dusty circumstellar material than that in typical Herbig stars.
	
	Much can be learned about the dust composition of the circumstellar material by analysing the mid-IR silicate spectral features. Such kind of studies {is} commonly performed among pre-main sequence stars \citep[e.g.,][]{vanBoekel2005,Juhasz2010,Maaskant2015}. Thus, analysing the mid-IR spectrum {can possibly} help to resolve the evolutionary state of HD 50138, and other unclassified B[e] stars. \citet{Fernandez2011} noted that the shape of the $10\ \mu$m spectral feature of HD 50138 indicates crystalline silicate grains (e.g., ortho-enstatite), but {they} did not make a detailed analysis. Mid-IR variability (including spectral variability and possible changes in the disc density and thermal structure) have also remained unexplored. Mid-IR interferometry may also provide clues for the existence of the putative binary companion. 
	
	

	In this paper we aim to address these questions by studying the circumstellar disc of HD 50138 with high spatial resolution mid-IR observations obtained with the VLTI/MIDI, presenting the first spatially resolved analysis of the mid-IR silicate spectral feature.
	The structure of the paper is as follows: In Sect. \ref{sec:obs} we describe the observations and the data reduction. In Sect. \ref{sec:res} we present the results of our interferometric analysis on the disc geometry, check for variability, 
	examine the mid-IR interferometric spectra, and search for signatures of binarity.
	In Sect. \ref{sec:dis} we analyse the properties of the circumstellar dust, and discuss if the new results could shed light on the evolutionary status of this enigmatic source. Finally, in Sect. \ref{sec:con} we summarize our results.
	\section{Observations and data reduction}
	\label{sec:obs}
	\subsection{Observations}
	The Mid-infrared Interferometric Instrument \citep[MIDI,][]{Leinert2003} was mounted on the Very Large Telescope Interferometer at ESO's Paranal Observatory, Chile. It was an interferometer, combining the light of either two 8.2 m Unit Telescopes (UTs) or two movable 1.8 m Auxiliary Telescopes (ATs). MIDI worked in the {7.5} to 13 $\mu$m spectral range, with $R \approx 30$ (prism), or $R \approx 230$ (grism) spectral resolution. 
	
\begin{figure}
		\begin{center}
			\includegraphics[width=\columnwidth]{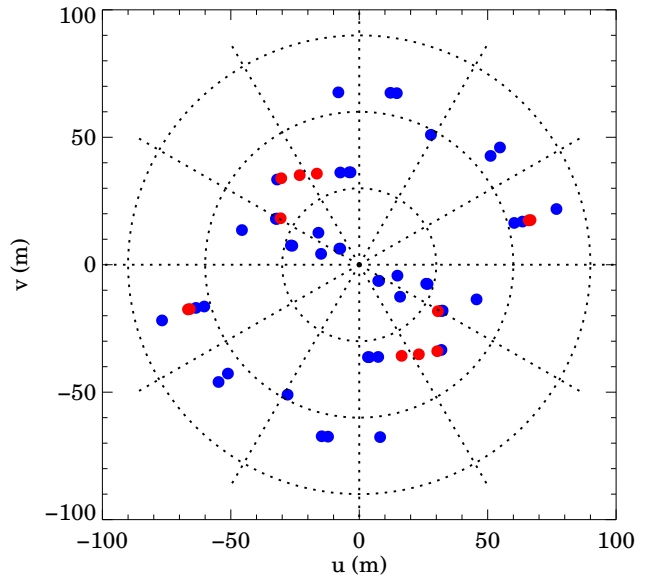}
			\caption{Distribution of the MIDI observations in the $uv$-plane. Red dots indicate bad quality measurements, not used for our further analysis.}
			\label{fig:uv}
		\end{center}
	\end{figure}

	Earlier MIDI observations of HD 50138 from 2007/08 and 2008/09 were already published by \citet{Fernandez2011}. We obtained additional new data in 2014/2015 under our ESO program 094.C-0629(B) (PI: P. \'Abrah\'am). The purpose of these measurements was to monitor the possible variability of the source. 
	All, earlier or new, observations were taken with ATs using the prism as the dispersing element. The distribution of the observations in the $uv$-plane are shown in Fig.~\ref{fig:uv}. Projected baseline lengths ($B_\mathrm{p}$) range from 9.7 m to 79.8 m. The position angles ($\phi_\mathrm{B}$) of the baselines are nearly evenly distributed. 
	
	\begin{table*}
		\caption{Overview of VLTI/MIDI interferometric observations of HD 50138, used in our analysis. $B_\mathrm{p}$ is the projected baseline length, $\phi_\mathrm{B}$ is the projected position angle of the baseline (measured {from} North through East), and $\diameter$ is the angular diameter of the calibrator star. $\phi_\mathrm{B}$ has a $180^{\circ}$ ambiguity. We use a convention where $\phi_\mathrm{B}$ is confined to the $\left(-90^{\circ},90^{\circ}\right]$ range.}
		\begin{center}
			\label{obs}
			\begin{tabular}{c c c c c c c c}
				\hline
				\hline
				\multicolumn{4}{c}{Target} & \multicolumn{3}{c}{Calibrator} \\
				\hline
				Date and time (UTC) & Configuration & $B_\mathrm{p}$ (m) & $\phi_B(^{\circ})$ & Name & {$\diameter$ (mas)} & Time (UTC)& Comment \\
				\hline
				2007-12-09T03:19:59	&	G1H0	&	68.9	&	-12.2	& HD 4128 & 5.19 & 02:25\\
				& & & & HD 48915 & 6.09 & 04:32\\
				2007-12-09T03:40:03	&	G1H0	&	68.5	&	-10.2	& HD 4128 & 5.19 & 02:25\\
				& & & & HD 48915 & 6.09& 04:32\\
				2007-12-12T05:09:38	&	G1D0	&	71.5	&	-50.0	& {HD 29139} & 20.40 & 02:35\\
				& & & & {HD 4128} & 5.19 & 03:39\\
				& & & & HD 48915 & 6.09& 04:47\\
				2007-12-12 06:00:52	&	G1H0	&	68.1	&	  6.9	& {HD 29139} & 20.40& 02:35\\
				& & & & {HD 4128} & 5.19 & 03:39\\
				& & & & HD 48915 & 6.09 & 04:47\\
				2007-12-13 03:10:16	&	G1D0	&	66.6	&	-50.1	& {HD 29139} & 20.40& 02:35\\
				& & & & HD 48915 & 6.09& 03:31\\
				2007-12-26 07:45:05	&	G1D0	&	58.1	&	-28.7	& HD 48915 & 6.09 & 07:03\\
				2008-11-10 05:43:04	&	E0H0	&	36.7	&	 60.4	& HD 48915 & 6.09& 05:18 & no measured total spectrum\\
				2008-11-10 05:47:01	&	E0H0	&	37.1	&	 61.0	& HD 48915 & 6.09& 05:18 & bad quality total spectrum\\
				2008-12-27 06:35:44	&	E0G0	&	15.5	&	 74.0	& HD 18884 & 12.27 &  02:57\\
				& & & & HD 48915 & 6.09 & 06:16\\
				2008-12-28 01:39:38	&	E0G0	&	10.0	    &	 50.8	& {HD 12929} & 6.90 & 00:36\\
				& & & & HD 48915 & 6.09 & 02:01\\
				& & & & HD 48915 & 6.09 & 02:23\\
				& & & & HD 48915 & 6.09 & 02:56\\
				2008-12-28 02:36:43	&	E0H0	&	36.9	&	 60.8	& {HD 12929} & 6.90& 00:36\\
				& & & & HD 48915 & 6.09 & 02:01\\
				& & & & HD 48915 & 6.09 & 02:23\\
				& & & & HD 48915 & 6.09 & 02:56\\
				2008-12-30T01:35:48	&	H0G0	&	20.3	&  51.7	& HD 12929 & 6.90& 00:22\\
				& & & & HD 48915 & 6.09 & 00:56\\
				& & & & HD 48915 & 6.09 & 00:59\\
				& & & & HD 48915 & 6.09 & 01:18\\
				2009-01-21T06:01:07	&	H0G0	&	27.6	&  74.1	& HD 31398 & 7.07 & 00:23 & no measured total spectrum\\
				& & & & HD 31398 & 7.07& 00:37\\
				& & & & HD 31398 & 7.07& 00:48\\
				& & & & HD 31398 & 7.07& 01:29\\
				& & & & HD 48329 & 4.62 & 02:22\\
				& & & & HD 48329 & 4.62 & 02:58\\
				2009-01-21T06:08:39	&	H0G0	&	27.1	&  74.0 & HD 31398 & 7.07& 00:23 & bad quality total spectrum\\
				& & & & HD 31398 & 7.07& 00:37\\
				& & & & HD 31398 & 7.07& 00:48\\
				& & & & HD 31398 & 7.07& 01:29\\
				& & & & HD 48329 & 4.62 & 02:22\\
				& & & & HD 48329 & 4.62 & 02:58\\
				2009-03-08T01:22:53	&	E0H0	&	47.6	&	 73.4	& {HD 48915}& 6.09   & 00:09\\
				& & & & HD 48915 & 6.09 & 01:41\\
				2014-10-14T06:31:05	&	C1A1	&	9.7	    &  49.5	& HD 12929 & 6.90 & 05:56 & no measured total spectrum\\
				& & & & {HD 32887} & 6.08 & 07:14\\
				& & & & {HD 32887} & 6.08 & 07:48\\
				2014-10-19T09:04:04	&	G1A1	&	79.8	&	-74.1	& HD 54605 & 3.60 & 08:48\\
				2014-11-15T04:52:53	&	G1A1	&	62.5	&	-74.8	& HD 54605 & 3.60 & 04:33 & bad quality total spectrum\\
				2014-11-17T03:57:34	&	G1I1	&	36.4	&	  5.4	& HD 54605 & 3.60 & 04:10\\
				2014-12-02T03:01:56	&	G1I1	&	36.5	&	  6.1	& {HD 12929} & 6.90 & 02:19\\
				& & & & {HD 48915}& 6.09 & 03:48\\
				& & & & HD 54605 & 3.60 & 03:20\\
				2014-12-02T04:02:18	&	G1A1	&	65.8	&	-75.1	& {HD 12929} & 6.90 & 02:19 & no measured total spectrum\\
				& & & & {HD 48915}& 6.09 & 03:48\\
				& & & & HD 54605 & 3.60 & 03:20\\
				\hline
			\end{tabular}
			
		\end{center}
	\end{table*}
	
	\begin{table*}
		\contcaption{}
		\begin{center}
			\begin{tabular}{c c c c c c c c}
				\hline
				\hline
				\multicolumn{4}{c}{Target} & \multicolumn{3}{c}{Calibrator} \\
				\hline
				Date and time (UTC) & Configuration & $B_\mathrm{p}$ (m) & $\phi_B(^{\circ})$ & Name & {$\diameter$ (mas)} & Time (UTC)& Comment \\
				\hline
				2014-12-02T04:19:41	&	G1A1	&	68.9	&	  -75.3	& 
				{HD 12929}& 6.90 & 02:19 & only total spectrum\\
				& & & & {HD 48915}& 6.09 & 03:48\\
				& & & & HD 54605 & 3.60 & 03:20\\
				2014-12-10T02:54:38	&	G1I1	&	36.9	&	 11.5	& HD 54605 & 3.60 & 02:40\\
				2015-01-09T04:56:11	&	G1I1	&	46.3	&	 43.7	& {HD 48915}& 6.09 & 04:23\\
				& & & & HD 54605 & 3.60 & 04:40\\
				\hline
			\end{tabular}
			
		\end{center}
	\end{table*}
	
	\subsection{Data reduction}
	\label{sec:red}
	We evaluated both the old and our new data, downloaded from the ESO public archive.
	The raw data are processed by the Expert Work Station (EWS) software \citep{chesneau2007,kohlerjaffe2008}. 
	Following \citet{Menu2015}, the calibration of the target was performed on a nightly basis by determining a transfer function. {Instead of taking the closest calibrator observation to determine the value of the transfer function, we perform a linear interpolation in time, by using different calibrators observed the same night} (Table~\ref{obs}). {Since the transfer function  is a time-dependent quantity, the advantage of our strategy is to correct for this time dependency.}  
	The transfer function includes the instrumental and sky responses. Then the total and the interferometric (correlated) spectra were calibrated separately. The final data products consist of the total spectra, correlated spectra, spectrally resolved visibilities and differential phases, all in the $7.5-13\ \mu$m spectral range. The uncertainties of the final spectra are dominated by systematic calibration errors of $10-20$ per cent. {The uncertainty of the correlated spectra are considerably smaller than that of the total spectra, because the background can be more efficiently subtracted from the interferometric measurements \citep[for more details, see][]{Menu2015}. The $9.4-10\ \mu$m wavelength range is affected by atmospheric ozone absorption, which makes the total spectrum calibration even more problematic. Thus, we do not use total spectrum data from this range in our analysis. The corresponding parts of correlated spectra, due to the more efficient background removal, are less affected by the ozone absorption, and are still usable.} The interferometric observations flagged as good quality by our data reduction, and useful for further analysis are listed in Table~\ref{obs}, and plotted in blue in Fig.~\ref{fig:uv}. {The table also lists the calibrators with their angular diameters taken from the calibrator database of Roy van Boekel, included in the MIDI data reduction package.} In a few cases the total spectrum measurements were unreliable (marked {as bad quality} in Table~\ref{obs}), and not used in the further analysis. {Some correlated spectra have no corresponding total spectrum observation, but we can still use them.}
	
	\section{Results}
	\label{sec:res}
	\subsection{Geometric modelling}
	\label{sec:modell}
	Interferometric observations can be used to characterize the geometry of circumstellar discs. \cite{Monnier2009} and \citet{Fernandez2011} have already determined the basic properties of the disc of HD 50138 at mid-IR wavelengths. They used various geometric models to determine the size, {the} inclination, and {the} position angle {($\phi$) of the major axis}. In the following, adding our new data from 2014/15, we re-evaluate the system morphology. First we use a simple Gaussian brightness distribution model to determine the orientation of the disc. Then we apply a semi-physical interferometric model to determine the size of the mid-IR emitting region and look for signs of variability. 
	
	
	For every observation we determined a characteristic size of the disc, defined as the half width half maximum (HWHM) of a Gaussian brightness distribution  fitted to the visibilities, according to the following formula:
	\begin{equation}
	\sigma=\frac{\sqrt{\ln2}}{\pi} \frac{\sqrt{-{\ln}V}}{B_\mathrm{p}/\lambda},
	\end{equation}
	where $\lambda$ is the wavelength, in this case $\lambda=10.7\ \mu$m, $V$ is the visibility at the selected wavelength, and $B_\mathrm{p}$ is the projected baseline length.
	
\begin{figure}
		\begin{center}
			\includegraphics[width=\columnwidth]{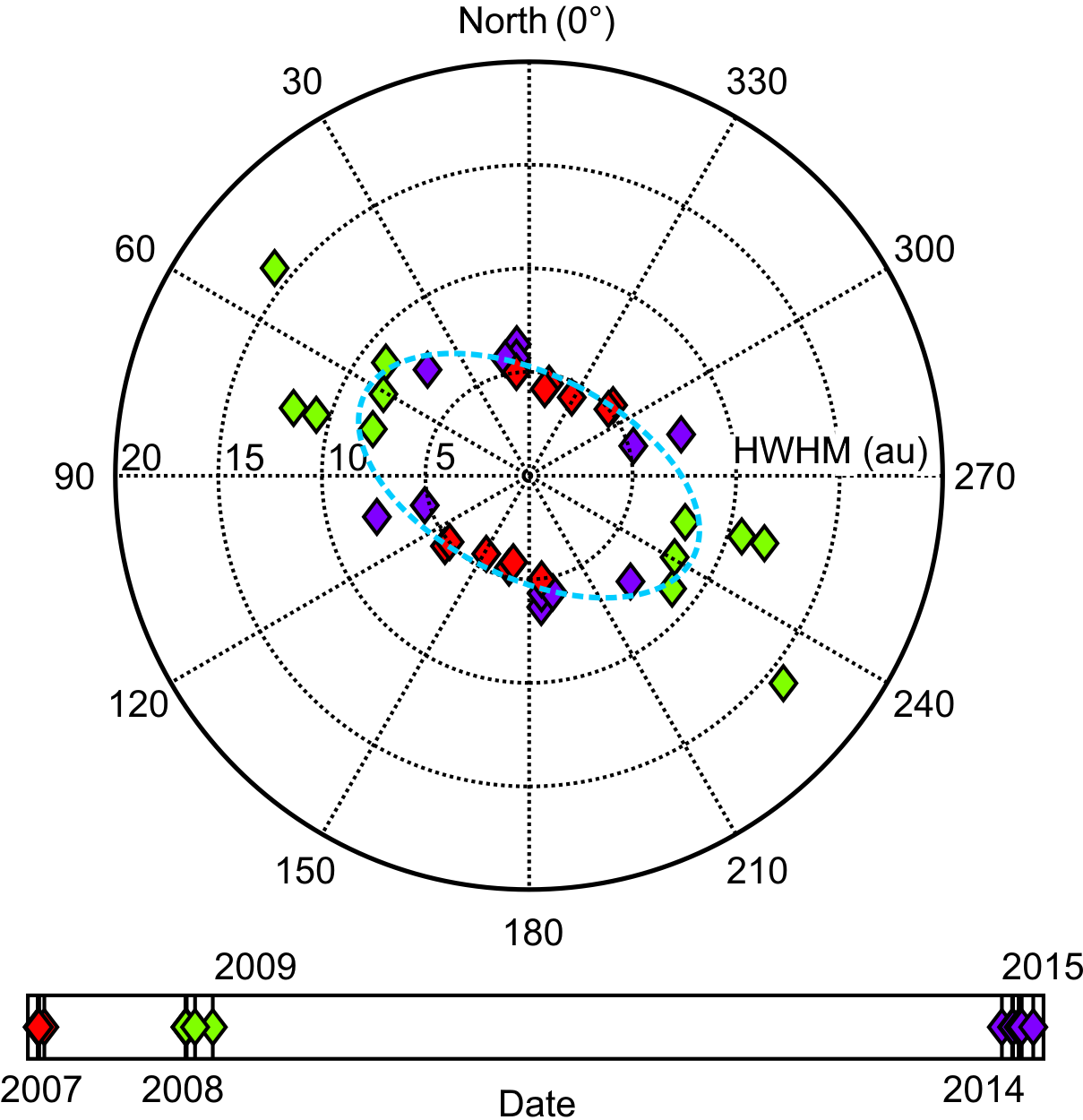}
			\caption{
				Gaussian sizes (HWHMs) derived from individual visibility values. The radial scale is in au. The angle is the position angle of the baselines (North is up). The fitted ellipse to the points is also shown. Points are colour-coded for observation date.}
			\label{fig:gauss}
		\end{center}
	\end{figure}

	The results are shown in Fig. \ref{fig:gauss}. We get sizes to different position angles. This enables us to measure the inclination of the disc. Therefore we fitted an ellipse to the resulting Gaussian sizes taking into account uncertainties with inverse squared error weighting. The fitted ellipse is plotted on Fig. \ref{fig:gauss}. From the plot it can be seen, that the data points from 2007 are close to the minor-axis of the projected image of the disc, and the data from 2008-2009 are aligned with the major-axis. Our new measurements, however, are at intermediate position angles.
	Disc inclination ($i$) is calculated simply by $i=\arccos \frac{b}{a}$, where $b$ and $a$ {are} the fitted semi-minor and semi-major axes, respectively. The resulting inclination is $56.6^{\circ\,+2.3^\circ}_{\ \, -1.1^\circ}$, and the position angle ($\phi$) {of the major axis} is $63.4^{\circ\,+1.5^\circ}_{\ \,-3.4^\circ}$ (measured from north to east), in agreement with the results of \citet{Fernandez2011}, although they used different interferometric model{s} to fit the data. {One of their models, which provided the best results, contained two elliptical Gaussian distributions, representing a compact and an extended component.} The semi-major axis of the fitted ellipse {in our fit}, i.e. the half-light radius (HWHM of the Gaussian model) of the disc is $8.3\pm 0.3$ au {($21.9\pm 0.8$ mas)}.  
	
	
	Pre-main sequence Herbig Ae/Be and T Tauri stars are known about their inherent brightness variability. The changing emission of the young star affects the irradiation of its disc, and the disc may show variability in its IR thermal emission, too. In order to outline any IR variability of HD 50138, we analysed all MIDI observations that cover a time span of seven years from 2007 to 2015. However, the direct comparison of interferometric observations taken at different baselines and position angles is not straightforward. We need to fit a geometric model which can predict the expected signal for each individual measurement. In the following we develop such a simple semi-physical model, and use it to compare the observations to search for temporal variability.
	
	
	The disc in our model is geometrically thin, circularly symmetric, and {it} is defined by an inner and an outer radius, manifesting a ring structure. We use the formerly derived inclination and position angle, and calculate effective baselines ($B_\mathrm{eff}$) to account for the disc orientation in the interferometric modelling. It is done with the re-projection of the projected baselines:
	\begin{equation}
	\begin{split}
	\alpha = \mathrm{atan2}\left(\sin\left(\phi_B-\phi\right),\cos i
	\cos\left(\phi_B-\phi\right)\right),
	\end{split}
	\end{equation}
	\begin{equation}
	\begin{split}
	B_{u, \mathrm{eff}} = B_\mathrm{p} \cos \alpha \cos \phi - B_\mathrm{p} \cos i \sin \alpha \sin \phi,
	\end{split}
	\end{equation}
	\begin{equation}
	\begin{split}
	B_{v, \mathrm{eff}} = B_\mathrm{p} \cos \alpha \sin \phi + B_\mathrm{p} \cos i \sin \alpha \cos \phi,
	\end{split}
	\end{equation}
	\begin{equation}
	B_{\mathrm{eff}} = \sqrt{B_{u, \mathrm{eff}}^2 + B_{v, \mathrm{eff}}^2}.
	\end{equation}
	
	{These equations describe the rotation and compression of the baseline reference frame. For more details we refer to \citet{Berger2007}.}
	Using effective baselines make{s} possible to use our data as if the disc {were} face on. If we suppose that the disc is circularly symmetric, then we do not need to take into account the position angle of measurements when comparing different epochs.
	
	In our model we assume that the disc radiates as a blackbody:
	
	\begin{table*}
		\caption{Results of the interferometric modelling.}
		\begin{center}
			\label{tab:res}
			\begin{tabular}{l l c c c c c }
				\hline
				\hline
				Parameter & Unit & \multicolumn{5}{c}{Wavelength ($\mu$m)}  \\
				&  & 8 & 9.2 & 10.7 & 11.3 & 13 \\
				\hline
				$q$ & & $0.62^{+0.03}_{-0.01}$&$0.64^{+0.02}_{-0.01}$&$0.70^{+0.03}_{-0.01}$&$0.72^{+0.03}_{-0.01}$&$0.80^{+0.04}_{-0.07}$\\
				
				$F_\mathrm{tot,\,\nu}$ & Jy & 46.6 & 67.9 &	66.0 & 64.5 & 39.8 \\
				$R_\mathrm{in}$ & au & $2.71^{+0.27}_{-0.05}$&$3.08^{+0.09}_{-0.01}$&$3.48^{+0.19}_{-0.02}$&$3.72^{+0.22}_{-0.09}$&$3.67^{+0.27}_{-0.57}$\\
				$R_\mathrm{in}$ & mas &  $7.19^{+0.70}_{-0.12}$ & $ 8.18^{+0.24}_{-0.03}$ & $ 9.24^{+0.50}_{-0.05}$ & $ 9.85^{+0.59}_{-0.23}$ & $ 9.74^{+0.71}_{-1.52}$\\
				\\
				$R_\mathrm{hl}$ & au & $8.55^{+0.18}_{-0.55}$&$9.48^{+0.01}_{-0.68}$&$9.22^{+0.10}_{-0.70}$&$9.18^{+0.31}_{-0.63}$&$8.24^{+1.18}_{-0.26}$
				\\
				$R_\mathrm{hl}$ & mas &  $ 22.68^{+0.47}_{-1.47}$ & $ 25.14^{+0.03}_{-1.79}$ & $ 24.44^{+0.28}_{-1.87}$ & $ 24.35^{+0.81}_{-1.69}$ & $ 21.87^{+3.14}_{-0.69}$ \\
				\hline
				
			\end{tabular}
			
		\end{center}
	\end{table*}
	
	\begin{figure}
		\begin{center}

			\begin{tikzpicture}
			\node[anchor=south west,inner sep=0] (image) at (0,0) {\includegraphics[width=\columnwidth]{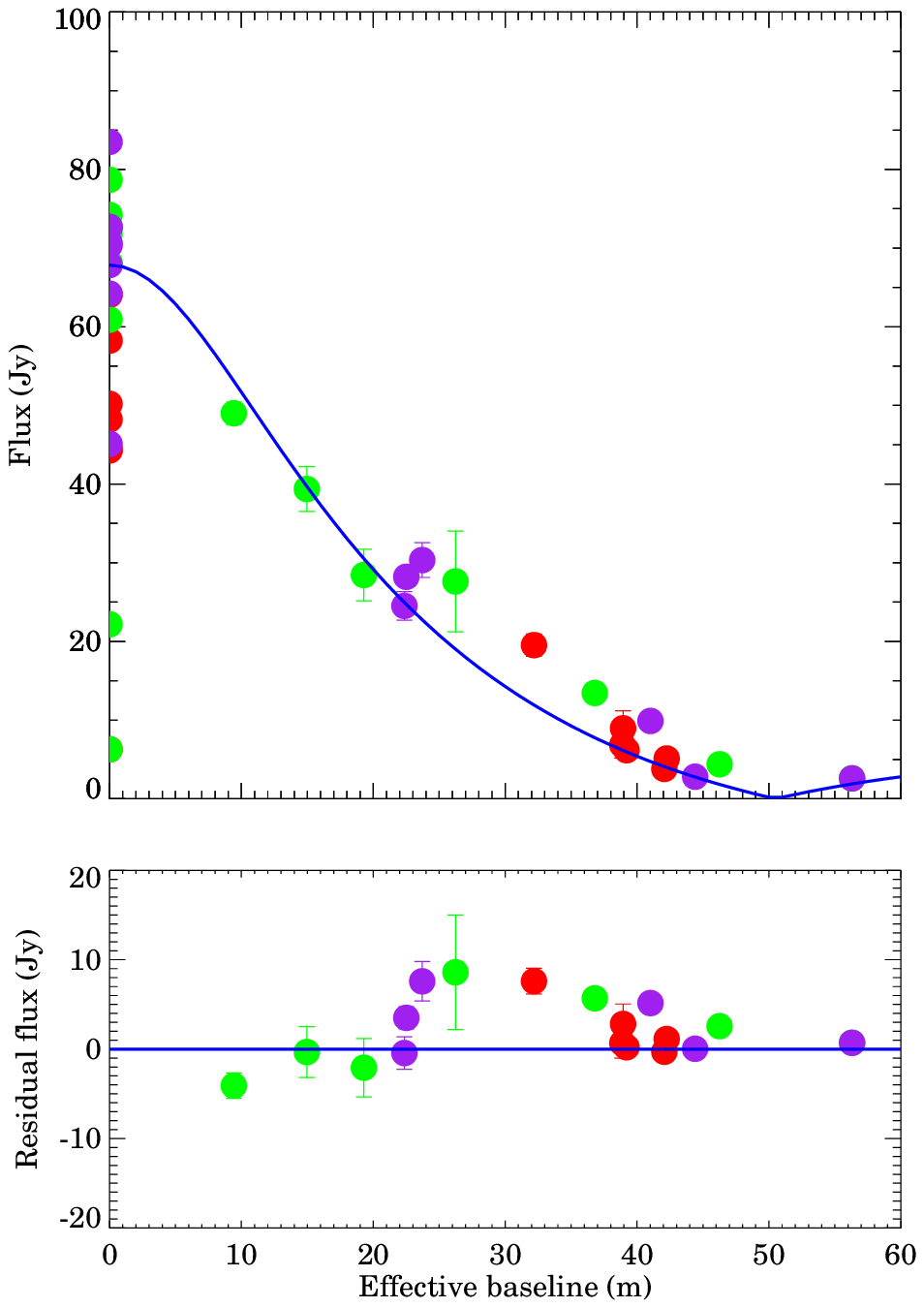}};
			\begin{scope}[x={(image.south east)},y={(image.north west)}]
			\node[anchor=south west,inner sep=0] (image) at (0.39,0.61) {\includegraphics[width=0.55\columnwidth]{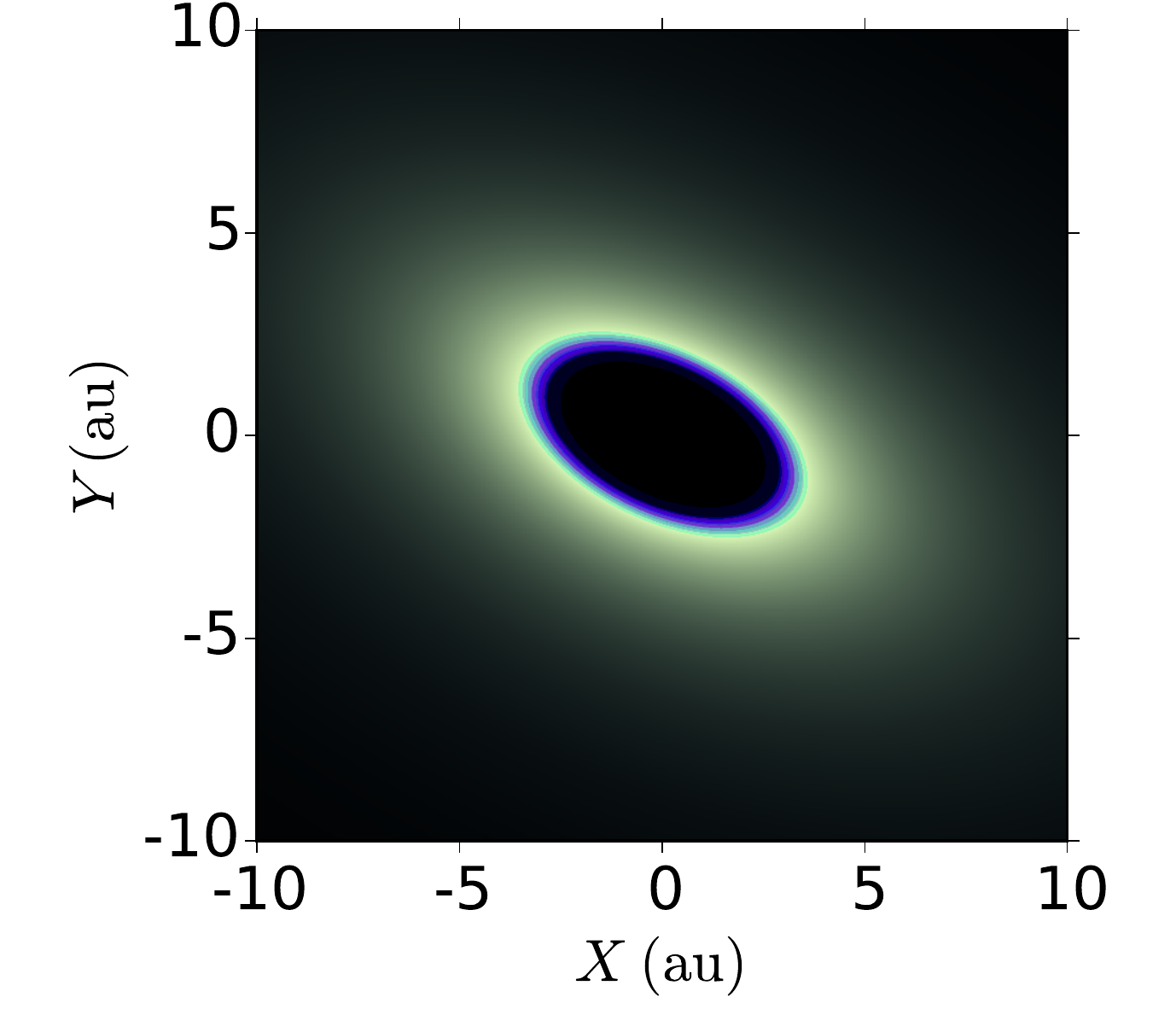}};
			\end{scope}
			\end{tikzpicture}
			
			\caption{Top: 10.7 $\mu$m correlated flux densities as a function of the effective baseline length, showing the result of the
				model fitting (blue curve). 10.7 $\mu$m total flux densities are also shown at zero baseline. Bottom: Residual plot. The symbols are colour coded for date of measurement, like in Fig.~\ref{fig:gauss}. Inset: false-colour image of the disc, based on the interferometric model. The colour channels correspond to $8-9.5\ \mu$m (blue), $10-11.3\ \mu$m (green), and $11.5-13\ \mu$m (red) wavelengths.}
			\label{fig:modell}
		\end{center}
	\end{figure}
	
	\begin{equation}
	I_\nu\left(r\right) = \tau_\nu B_\nu\left(T\left(r\right)\right),
	\end{equation} 
	\noindent where $I_\nu$ is the wavelength dependent intensity, $\tau_\nu$ is the opacity, $B_\nu$ is the Planck-function, and $T$ is the temperature. The dependency of the temperature on the radius ($r$) is a power-law:
	\begin{equation}
	T\left(r\right) = T_\mathrm{in}\left(\frac{r}{R_\mathrm{in}}\right)^{-q}.
	\end{equation} 
	\noindent where ${R_\mathrm{in}}$ is the inner radius, $T_\mathrm{in}$ is the temperature at ${R_\mathrm{in}}$, 
	and $q$ defines the temperature gradient. {Following \cite{Menu2015}, we derive the relation between $R_\mathrm{in}$ and $T_\mathrm{in}$ as} 
	\begin{equation}
	R_{\mathrm{in}} = \left(\frac{L_\star}{4\pi\sigma T_\mathrm{in}^4} \right)^{1/2}
	\label{eq:rin}
	\end{equation} 
	{where $L_\star$ is the luminosity of the star. For $L_\star$ we use the value by \citet{Ellerbroek2015}, but recalculating it for the new Gaia DR2 distance measurement: $L_\star = \left(680 \pm 230\right)\ L_\odot$.  }
	The outer radius is fixed at $300$~au where the mid-IR emission is negligible.
	The total flux density, {which is the integrated emission of the source over the whole disc} ($F_\mathrm{tot,\nu}$), and the correlated flux density ($F_\mathrm{corr,\nu}$) are derived from the following formulas:
	
	\begin{equation}
	F_\mathrm{tot,\nu} = \int_{R_\mathrm{in}}^{R_\mathrm{out}} 2 \pi r \tau_\nu B_\nu\left(T\left(r\right)\right) \mathrm{d}r.
	\end{equation} 
	\begin{equation}
	F_\mathrm{corr,\nu}\left(B_{\mathrm{eff}}\right) = 
	F_\mathrm{tot,\nu} 
	\frac
	{\int_{R_\mathrm{in}}^{R_\mathrm{out}} r B_\nu\left(T\left(r\right)\right) J_0\left(2\pi r B_{\mathrm{eff}} / \left( \lambda d \right) \right) \mathrm{d}r }
	{\int_{R_\mathrm{in}}^{R_\mathrm{out}} r B_\nu\left(T\left(r\right)\right)  \mathrm{d}r }.
	\end{equation}
	\noindent where $J_0$ is the zero-order Bessel-function, and $d$ is the distance of the star. 
	The half-light-radius ($R_\mathrm{hl}$) of the disc is {defined as the radius at which the integrated intensity is half of the total flux density:}
	\begin{equation}
	\frac{F_\mathrm{tot,\nu}}{2} = 
	\int_{R_\mathrm{in}}^{R_\mathrm{hl}} 2 \pi r I_\nu\left(r\right) \mathrm{d}r.
	\end{equation}
	
	Based on our model, we can compare the MIDI data points and check for potential variability, accounting for the different effective baselines and position angles of the individual measurements. Figure~\ref{fig:modell} shows the data, overplotted with the best-fitting model. {The large scatter in the total flux data (at zero baseline) is mainly due to the larger uncertainty in the total spectrum calibration, compared to the interferometric measurements (Sect.~\ref{sec:red}).} In the fitting process we varied two free parameters, $q$ and $R_\mathrm{in}$. The best-fit parameters are computed in a grid, by maximizing the log-likelihood. Table~\ref{tab:res} presents the results for five selected wavelengths. 
	
	
	
	We have got $2.7-3.7$~au {($7.2-9.9$~mas)} for the inner radius, with an increasing trend with wavelength. We can compare this result with the dust sublimation radius {($R_{\mathrm{sub}}$), which can be computed from Eq.~\ref{eq:rin}, by substituting $T_\mathrm{in} = T_\mathrm{sub} = 1500$~K for the sublimation temperature \citep{Dullemond2010}.}  
	The corresponding sublimation radius is $1.8 \pm 0.3$~au {($4.7 \pm 0.8$~mas)}. This is smaller than the observed inner radius, suggesting the presence of a hole in the dust disc. {For a certain inner radius we can compute an equilibrium temperature using Eq.~\ref{eq:rin}.} The equilibrium temperatures at the inner disc edge are in the range of $1050-1200$~K. The mid-IR measurements trace the distribution of the dust. However, the detected hole is not empty, as gas was detected within this radius {\citep{Ellerbroek2015,Koutoulaki2018}}. This is the first detection of an inner {dusty} disc hole (which is larger than the sublimation radius) in the dusty disc of a B[e] star to our knowledge. We note, however, that combined multi-wavelength data-sets are needed to unambiguously determine the exact parameters of the inner {dusty} disc hole \citep[as an example, we refer to][]{Menu2014}. 
	
	{Similar inner disc gaps are frequently observed in young intermediate-mass systems, in particular in transitional discs \citep{Espaillat2014}, but were also detected with MIDI around other Herbig Ae/Be stars \citep{Menu2015}. There are several clearing mechanisms proposed, like grain growth, photoevaporation, and dynamical clearing by a (sub-)stellar companion \citep{Espaillat2014}. \citet{Kluska2016} reported a moving asymmetry in the disc of HD 50138, which could be fitted with a circular trajectory with a radius of $\sim$2~au. However, they could not directly link the asymmetry to the presence of a companion, rather they hypothesized that it may be a spiral arm or other disc asymmetry. We note that spiral arms in circumstellar discs can be induced by companions or planets \citep[e.g.,][]{Bae2018,Boehler2018,Dong2018,Muller2018,Wagner2018}. Such a scenario can be compatible with the observed features of the disc around HD 50138.} 
	
	Near-IR interferometric observations of HD 50138 \citep[e.g.,][]{Lazareff2017,Koutoulaki2018} yielded a smaller size {($0.7-1.1$~au)} for the continuum-emitting region. Such deviations can result from the existence of an additional continuum-emitting component inside the sublimation radius, usually attributed to refractory dust or to hot gas by some authors \citep[e.g.,][]{Benisty2010}. {Alternatively, an optically thick inner disc applied to disc models of Be stars, similar to the pseudo-photospheres in luminous blue variables, can also provide an explanation \citep{Vieira2015}.} The nature of this component however, is not very well constrained.

	From our model we determine a range of $8.2-9.5$~au {($ 21.9-25.1$~mas)} for the half-light radius as a function of wavelength. We found that the half-light radius at wavelengths corresponding to the continuum ($R_\mathrm{hl} = 8.2-8.6$~au) is smaller than at wavelengths corresponding to the silicate feature ($R_\mathrm{hl} = 9.2-9.5$~au). This means that the optically thick continuum is confined to a smaller area than the optically thin silicate emission.
	
	We checked the residual distribution of the data points around the best fit model line in Fig.~\ref{fig:modell}. We mark the epochs of the observations with colour. In the $25\ \mathrm{m} < B_\mathrm{eff} < 40$~m range some data points are significantly above the model, 
	possibly indicating fine structures not accounted by our simple geometry. However, we found no trend in the residuals with time, 
	thus excluding any significant variability in the mid-IR {morphology} of the disc over the observed seven year time range. {Possible variations in the total flux data will be discussed in Sect.~\ref{sec:var}.}

	\subsection{The 10~$\mu$m silicate feature}
	\label{sec:sil}
	
	\begin{figure}
		\begin{center}
			\includegraphics[width=1.0\columnwidth]{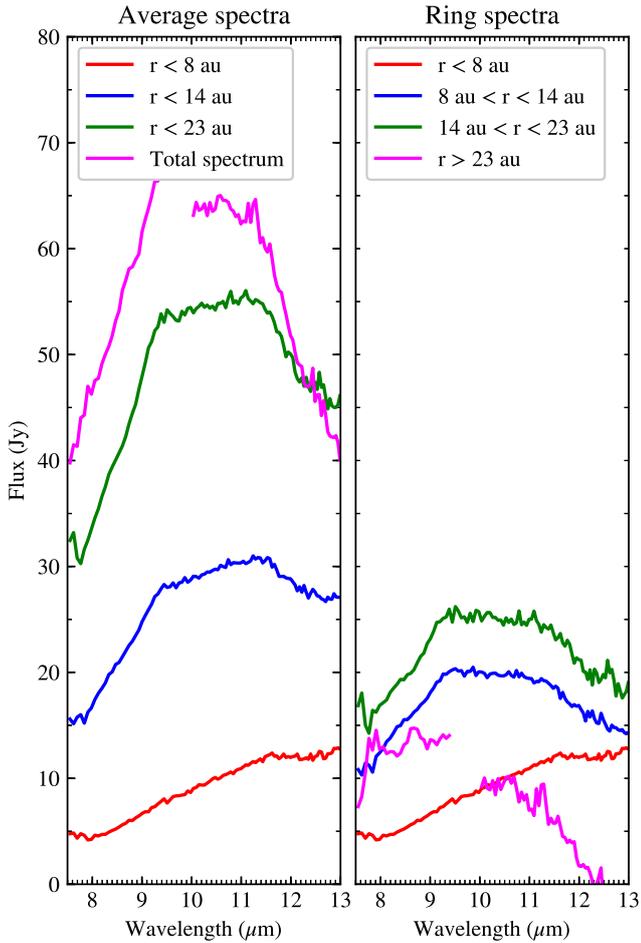}
			\caption{Averaged spectra of various radial disc regions, derived from the correlated and total spectra. Left: integrated emission within given radii. Right: differential spectra from ring-shaped regions. }
			\label{fig:spec_all}
		\end{center}
	\end{figure}
	
	There are several notable spectral features observable in circumstellar discs in the N-band, such as the features of silica, various silicates, and polycyclic aromatic hydrocarbons (PAHs). The MIDI spectra of HD 50138 have sufficient spectral resolution ($R \approx 30$) and signal-to-noise ratio to study these features. The interferometric observations enable us to study any spatial variability in the spectra. With our time coverage of 7 years, we can also look for potential temporal changes in the spectra. 
	
	We sorted the observed correlated spectra into several groups, based on their inclination-corrected angular resolution, calculated from the effective baseline lengths. Angular resolution is measured as the half width at half maximum (HWHM) of a fringe. We defined three bins in resolution, corresponding to different regions of the disc: $3-8$ au ({$8-21$~mas,} innermost disc), $8-14$ au ({$21-37$~mas,} inner disc), and $\sim$23~au ({$\sim$61~mas,} intermediate disc). As a fourth bin, we included the total spectra, which contain emission from the full disc. 
	Since within the individual groups there was no obvious dependence of the spectral shape on the position angle, we averaged the spectra in each bin. The resulting spectra correspond to all integrated emission within a given disc radius determined by the resolution. The results are plotted in the left panel of Fig.~\ref{fig:spec_all}. In order to explore the differences among the radial disc regions, in {the right panel of} Fig.~\ref{fig:spec_all} we present differential spectra between the averaged spectra, which represent  emission of ring-shaped regions in the disc.
	
	The averaged spectra, {both the total and correlated ones,} exhibit a broad emission feature at $10\ \mu$m, characteristic of the emission of small silicate grains. The correlated spectra show clear {edges} in the spectral  feature at $9.2$ and $11.3\ \mu$m. These turning points seem also to be present in the noisier total spectra. The profiles, however, show significant variations in how strong are these turning points.
	The innermost disc spectrum ($r < 8$~au) is relatively flat, nevertheless it exhibits a definite change in its slope at $11.3\ \mu$m. While the intermediate regions show both {edges} with a characteristic plateau between $9.2$ and $11.3\ \mu$m. The $11.3\ \mu$m edge in the total disc spectrum is less conspicuous. The amplitude of the feature gets smaller in the inner and innermost disc regions. 
	
	
	

	\subsection{Binarity}
	
	\begin{figure}
		\begin{center}
			\includegraphics[width=1.0\columnwidth]{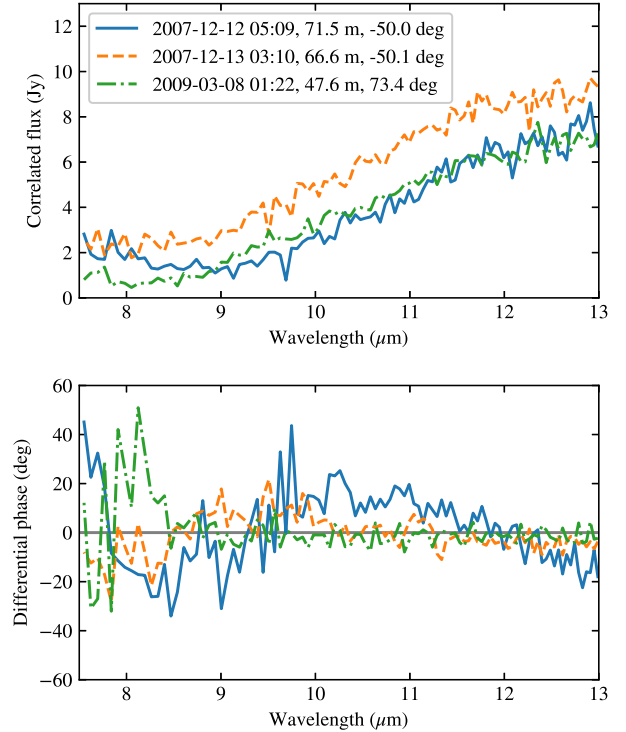}
			\caption{{Top: Correlated spectra which show spectral shapes resembling to binary modulation.  Bottom: The corresponding differential phase curves. The labels list the observation date, projected baseline length and position angle.} }
			\label{fig:spec_phi}
		\end{center}
	\end{figure}
	
	Following previous near-IR interferometric studies that observed disc asymmetries \citep[e.g.,][]{Kluska2016,Koutoulaki2018}, we examined the MIDI data to find indications for binarity. The binary signal in the spectrally resolved correlated spectra and differential phases usually appears as a sinusoidal modulation. Some of the highest resolution correlated spectra of HD 50138 show patterns, which might be interpreted as a binary modulation. In Fig.~\ref{fig:spec_phi} we show three spectra of this kind (top panel). As the correlated spectra also exhibit silicate emission, it is difficult to disentangle the potential sinusoidal binary signal from the spectral feature. A companion or an asymmetric disc structure would also cause a modulation in the differential phases, plotted on the bottom panel. The differential phase curves show some deviations from the flat zero degree line, but the signal is generally too weak and noisy to be able to confirm binarity or more generally an asymmetric disc structure. MIDI is sensitive to binaries with separations larger than $\sim$50~mas, with mid-IR flux ratios larger than $\sim$$0.1$ \citep{varga2018}. Since we do not see clear indication for binarity from these data, the companion is either closer to the star than this limit, or is too faint in mid-IR, or both. Furthermore, the possibility that there is no companion cannot be discarded.

	\section{Discussion}
	\label{sec:dis}
	
	\subsection{Dust properties}
	
	Cosmic dust grains are predominantly formed in the stellar wind of AGB stars, then distributed in the interstellar space. {\citet{Miroshnichenko2007} discusses the origins of dust around different classes of B[e] stars: Two groups can be defined regarding the dust formation time, one is where the formation happened during earlier evolutionary phases, the other is where recent or ongoing dust formation can be observed. Herbig B[e] stars, where the dust has an interstellar origin (i.e. it is produced by previous generations of stars), and compact planetary nebulae B[e]-type stars, where the dust is left over from the previous AGB state, are examples for the first case. All other classes, i.e symbiotic, supergiant, and unclassified B[e] stars have freshly produced dust around them. In the latter two classes the companion is possibly responsible for the mass loss, and the dust is formed in a dense radiation-driven stellar wind, where self-shielded parts (from UV radiation) are the cradles for dust condensation. \citet{Miroshnichenko2007} also suggests that the dust in unclassified B[e] stars should form around the binary, as a consequence of a recent or ongoing rapid mass transfer phase. However, the same author claims that only $30\%$ of unclassified B[e] stars are suspected or recognized binaries. Thus, a single star scenario with dust formation in the circumstellar disc cannot be discarded.
	\citet{Fuente2015} presented an alternative hypothesis, that a stellar merger can also explain the formation of large amounts of circumstellar dust, thus unclassified B[e] stars could be the products of a recent merger. From the observational side, \citet{Miroshnichenko2011} studied the mid-IR spectra of 25 unclassified B[e] stars, and they detected silicate emission features in most observed objects, and PAH emission in some.} 
	
	{Discs around Herbig B[e] stars are formed by the collapse of the protostellar cloud, which contains interstellar dust.} The majority of {these} grains are small, sub-micron sized amorphous silicate particles \citep{Kemper2004}. In circumstellar discs, however, grains can be reprocessed through crystallization and grain growth. There are several crystal formation mechanisms, like thermal annealing in the inner disc
	above a temperature of $\sim$1000~K;
	re-condensation after evaporation; or heating by shock waves \citep{fabian2000,harker2002,gail2004}. {\cite{Gail1999} found that crystalline olivine can even directly condense out from the stellar wind of oxygen-rich mass-losing stars.} Crystalline grains are found throughout the discs, indicating that if crystalline dust formed close to the star, it had to be transported to the outer disc. Radial transport mechanisms, such as disc/stellar winds are proposed for protoplanetary discs \citep[e.g.,][]{Gail2001,gail2004}. Alternatively, shock waves due to gravitational instabilities, or episodic outbursts can induce crystal formation at a greater distance from the central source. 
	
	Our MIDI spectra exhibit characteristic silicate emission profiles, with two conspicuous {edges} at $9.2$ and $11.3\ \mu$m, and with a variety of the relative strength of the feature above the continuum (Sect.~\ref{sec:sil}, Fig.~\ref{fig:spec_all}). Such {features} may imply the presence of crystalline silicates of different types, but may also be related to grain growth \citep{vanBoekel2003,Przygodda2003}. {Additionally, the replenishment of dust at the disc surface by large grains, coming from the midplane or inner regions, may also take place here \citep[e.g.,][]{Ciesla2007,Turner2010}.} Thus{,} it is hard to determine the origin of the spectral features unambiguously. In the following{,} we will try to interpret our MIDI results in {different} scenarios, and draw conclusions about the radial dust composition of the inner circumstellar disc.
	
	In order to explore if the spectral shape could be due to crystalline silicates, we compare the features present in the spectra of HD 50138 with various laboratory spectra of silicate dust grains from \citet{vanBoekel2005} (Fig.~7 therein). 
	To determine the exact dust composition, we would need very high signal-noise ratio \citep{Juhasz2010}. The MIDI spectra are less adequate for such analysis, nevertheless we can decide whether amorphous or crystalline dust is the dominant component, and we can identify the most likely mineral types. Fig.~7 of \citet{vanBoekel2005} suggests that the shoulder at $11.3\ \mu$m may be attributed to crystalline forsterite particles. While PAHs may also exhibit a peak at this wavelength, the fact that the feature is present in the correlated spectra makes a PAH interpretation unlikely. The emission of these carbonaceous particles is expected to be distributed smoothly over {a very large spatial region, due to their transient heating \citep{Allamandola1985}}, {compared to the more compact 10~$\mu$m silicate emission.} Thus, it would be {mostly} filtered out by the interferometer{, as such large structures become fully resolved at the baselines we used. Consequently the interferometric signal of the PAHs is very low at usual VLTI configurations, compared to the warm, silicate emitting circumstellar material. PAH features could be seen in the MIDI total spectra, but we find no indication for it in the case of HD 50138. In the correlated spectra} we see a shoulder around $9.2\ \mu$m. This wavelength is somewhat shorter than that of the strong forsterite peak at $\sim$10 $\mu$m, thus we propose that the $9.2\ \mu$m feature in our spectra is more likely to be either enstatite or silica {\citep[see Fig.~7 from][]{vanBoekel2005}}. A contribution from small amorphous particles would follow a sharp, triangular shape emission feature peaked at about $9.8\ \mu$m, which may contribute to the outer disc spectrum.
	
	In the framework of the crystalline scenario, one would conclude that the innermost part of the HD 50138 disc (within $8$~au) is mainly populated by forsterite grains, while at intermediate radii $8-23$~au) also enstatite appears with comparable amplitude. {This radial distribution might suggest that in the innermost disc the temperature is too high for enstatite, although from our modelling we expect $T \approx 600$~K at 8~au, indicating that both species can exist in the $<8$~au region (for dust condensation temperatures we refer to \citealp{Grossman1972} and \citealp{Tielens1990}). Maybe the fraction of enstatite in the innermost region is too low to detect it in the MIDI spectra.} The outer part of the disc ($r>23$ au) lacks the $11.3\ \mu$m {edge}, but exhibits a {broad} peak {between 8 and $9~\mu$m}. 
	{The shape of this feature does not well match with the templates of \cite{vanBoekel2005}, but it may correspond to silica or small amorphous (pyroxene) particles.}
	Considering, however, the relatively large radii, and that crystallization mostly take place in the hot inner disc, the latter explanation is more appealing. A similar trend, that the level of crystallinity is higher toward the centre and lower in the outer disc was observed by \citet{vanBoekel2004}.
	
\begin{figure}
		\begin{center}
			\includegraphics[width=\columnwidth]{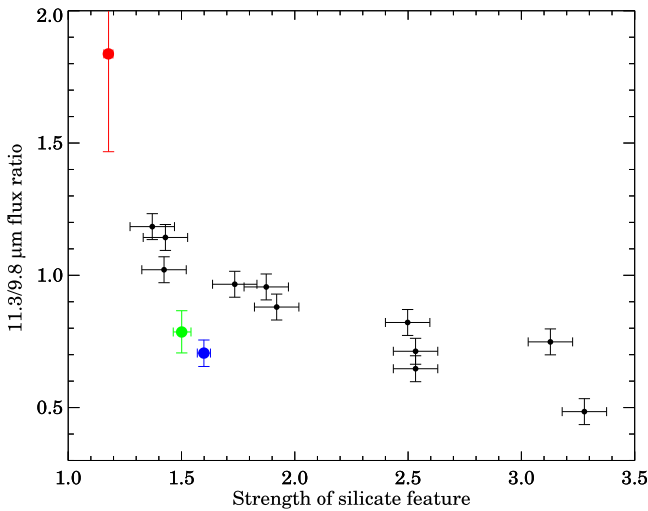}
			\caption{$11.3/9.8\ \mu$m flux ratios of the different disc regions in HD 50138 as a function of the strength of the normalized silicate emission. The coloured symbols correspond to the spectra in Fig.~\ref{fig:spec_all} right panel. For comparison, results for a sample of Herbig Ae/Be stars from \citet{vanBoekel2003} are overplotted (black dots).}
			\label{fig:boekelplot}
		\end{center}
	\end{figure}

	
	In the presence of a population of larger, micrometer-sized silicate grains, the profile of the 10~$\mu$m peak will exhibit a characteristic trapezoidal shape \citep{Bouwman2003}. Studying a sample of young intermediate-mass {Herbig Ae/Be} stars, \citet{vanBoekel2003} found a linear relationship between the ratio of the 11.3 to the 9.8 $\mu$m fluxes of the continuum-subtracted spectra and the strength of the silicate feature (Fig.~\ref{fig:boekelplot}). The authors interpreted this relationship as {evidence for the removal of small ($\sim0.1~\mu$m) grains from the disc surface while large ($1-2~\mu$m) grains remain, and for the increased presence of crystalline silicates.} In HD 50138 the spectral shapes at intermediate disc radii ($8<r<23$~au) are trapezoidal. In order to check whether these spectra could be interpreted in terms of grain growth, following \cite{vanBoekel2003} we computed the normalized spectra with the following formula:
	\begin{equation}
	F_{\nu\mathrm{,\,norm}} = \left(F_{\nu} - F_{\nu\mathrm{,\,cont}}\right) / \langle F_{\nu\mathrm{,\,cont}} \rangle + 1 
	\end{equation}
	\noindent where $F_{\nu\mathrm{,\,norm}}$ is the wavelength dependent normalized flux,  $F_{\nu}$ is the flux, and $F_{\nu\mathrm{,\,cont}}$ is the continuum which we determined from a linear fit to the data points between $7.5$ and $8.0\ \mu$m and between $12.5$ and $13\ \mu$m. The strength of the feature is then determined as the maximum of $F_{\nu\mathrm{,\,norm}}$. The flux ratio is derived from the continuum-subtracted spectra. The results, computed for the different ring-shaped disc regions of Fig. \ref{fig:spec_all}, are overplotted in Fig.~\ref{fig:boekelplot}. {In case of the outer disc spectrum ($r>23$~au) we could not get a reliable estimation, because of the large uncertainty of the linear continuum fit. Thus we do not show the corresponding data point in the figure.} The data points representative of {three} disc regions outline a trend, that is significantly steeper than the relationship of the Herbig Ae/Be stars. We suggest that the spectral shapes in HD 50138 are more consistent with the presence of crystalline grains in the disc than with grain growth.
	
	
	Now we try to characterize the expected differences between the dust content of pre-main-sequence B[e] and unclassified B[e] stars in order to shed light on the origin of dust around HD 50138. In pre-main-sequence stars the dust mainly originates from an existing interstellar cloud, while in unclassified B[e] stars it is supposed to form in situ around the star, in an inside-out fashion. This should result in observable differences between these classes. Therefore the dusty discs around older B[e] stars are expected to be compact, and the freshly condensed dust grains should be very small \citep{Lee2016}. In the case of HD 50138 we showed that the mid-IR emitting region has a size of $R_\mathrm{hl} \sim 9$~au. We can compare this size to the results of \citet{Menu2015}, who studied the size distribution of discs around pre-main-sequence stars. We find that the mid-IR size of HD 50138 is similar to the size of discs around Herbig Be stars with similar luminosities (which are in the range of $\sim3-10$~au). We note, however, that a really compact appearance for older B[e] stars is expected at longer wavelengths, due to the lack of cold dust in these systems. \citet{Lee2016} found that the fall-off of the SEDs of HD 50138 and HD 45677 toward submillimeter wavelengths suggests a lack of cold grains, thus more compact dusty envelopes than those around Herbig stars. \citet{Sandell2018} confirmed that there is very little cold dust in the disc of HD 50138. Furthermore, they found that $^{13}$CO in the disc is enriched by more than a factor of five, which is uncommon for young stellar objects, supporting that HD 50138 is not a Herbig star.

	{The other expectation for dust around older B[e] stars is the presence of freshly produced grains. In the case of HD 50138 we clearly see the signs of dust evolution, with the more processed dust located closer to the star. This seems to contradict the inside-out dust formation scenario, and a similar trend were observed in discs around pre-main sequence stars \citep[e.g.,][]{vanBoekel2004,varga2018}. This argument might seem to support the interstellar origin of the dust around HD 50138. We note, however, that crystalline silicates are frequently found around evolved stars (AGB stars and red supergiants) \citep[e.g.,][]{Waters1996,Speck2000,Liu2017}, and many aspects of the dust condensation sequence and crystallization are still unclear.  A possible scenario, which was proposed for AGB and red supergiant stars \citep{Waters1996}, is that in dense parts of the stellar wind the dust can stay warm long enough to allow formation of crystals.}

	\subsection{Variability}
	\label{sec:var}
	
	\begin{figure}
		\begin{center}
			\includegraphics[width=\columnwidth]{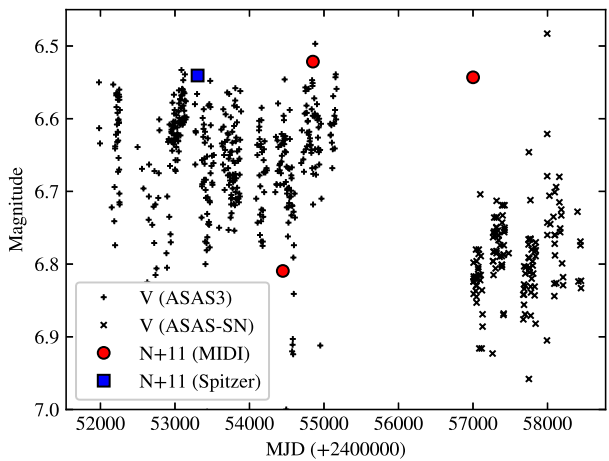}
			\caption{Optical V band (black symbols) and mid-IR N band (blue and red points) light curves of HD 50138. The mid-IR points are shifted by 11 mag. The MJDs for the MIDI data points are 54447.70, 54851.41, and 57000.90. }
			\label{fig:lc}
		\end{center}
	\end{figure}
	
	{HD 50138, as other B[e] stars, shows permitted and forbidden emission lines in its spectrum, which traces the gas content. The dust is observed by its IR excess and the silicate spectral feature. A notable result of our geometric modelling, that we do not detect significant changes in the structure of the N band emitting disc region (Sect. \ref{sec:modell}). This is different from the results on both the optical spectral line variations \citep{Jerabkova2016} and the changes in the near-IR morphology of the disc \citep{Kluska2016}. The location of the optical and near-IR variable features is likely the central area of the system, inside $r \approx 2$~au, which is the radius of the rotating H band asymmetry. From our modelling we find that the N band emitting region begins at $R_\mathrm{in} \approx 2.7-3.7$~au. Thus the more dynamic innermost and less variable outer part of the system may be spatially separated. 
		We note, however, that the temporal coverage of our observations are sparse, and thus not well suited for variability analysis. Additionally, the time scales we can probe (typically a few weeks to years) are either too long or too short compared to most periods of optical spectral variations \citep{Pogodin1997,Fernandes2012,Jerabkova2016}.}
	
	{Apart from the morphology, we can also look for variations in the total N band flux. We divided our observations into three bins, representing the measurements from 2007, 2008/2009, and 2014/2015. We then calculated a median total spectrum in each bin. We found that the average N band flux was 30 per cent lower in 2007 than later, which indicates real time variability. Comparing this finding with the optical line monitoring of \citet{Jerabkova2016}, we do not observe correlations between the long-term trends of the optical and mid-IR variations. }
	
	{In Fig. \ref{fig:lc} we  compare the N band flux variations to optical photometric data from the ASAS3 \citep{Pojmansk1997} and ASAS-SN \citep{Shappee2014,Jayasinghe2018} databases. We plot an additional N band data point calculated from a Spitzer Infrared Spectrograph observation\footnote{{The spectrum was downloaded from the Combined Atlas of Sources with Spitzer IRS spectra \citep{Lebouteiller2015}. AORKey: 11003648.}} from 2004-10-26. The optical measurements by the ASAS3 outline a trend showing a slight dip around $\mathrm{MJD\ (+2400000)}=54000$, which may be correlated with the first three N data points. The average magnitude level of the ASAS-SN data is $\sim0.15$~mag lower than that of the ASAS3 data, however, the last MIDI flux is still high. Therefore the correlation may not hold for an extended period of time.
		\citet{Fernandes2012} analysed the optical line profile variability of HD 50138 and also confirmed night-to-night variations. They pointed out that strong short-term changes are linked to the lines formed in the upper layers of the stellar atmosphere, while lines originating in the circumstellar environment (e.g., }[\ion{O}{i}]{) are not variable over daily time scales.
		HD 50138 appears to be very complex, where a range of variable phenomena can take place, e.g., stellar pulsation, shell phases, stellar wind, orbiting companion, rotating disc asymmetries. Systematic multi-wavelength monitoring from the optical to the mid-IR would be required to physically connect these processes happening on different time scales in different regions of the circumstellar environment. } 
	
	\subsection{Evolutionary state}
	
	Despite numerous efforts, the evolutionary state of unclassified B[e] stars is still uncertain. In case of HD 50138 there is a slight preference in the literature for {either late main-sequence or} post-main sequence state \citep[e.g.,][]{Fernandes2009,Ellerbroek2015}. {\citet{Fernandes2012} observed short-term optical line profile variability indicating stellar pulsations similar to pulsating Be stars. They also suggested that the object is at the end, or just evolving off the main sequence.} \citet{Kluska2016} placed the star on the Hertzsprung-Russell diagram, and fitted isochrones for a young and an old scenario. They got ages of $0.3-0.5$~Myr for the young, and $100-200$~Myr for the old scenario. Both scenarios are consistent with the observations. 
	
	Here we compare our observations of HD 50138 with publicly available spectra of other B[e] stars from the MIDI archive to find possible correlations between the spectral shape and the known evolutionary state in different subgroups. We expect such kind of correlations as the physical conditions for the disc formation and evolution around supergiants and around pre-main sequence stars can be very different. Our sample consists of supergiants, young Herbig Ae/Be stars {(including three Herbig Be stars)}, and unclassified B[e] stars. The MIDI data were consistently reduced and calibrated using the methods described in Sect.~\ref{sec:red}. Looking at the $8-13\ \mu$m total spectra, we can distinguish two main groups in the sample: one group showing silicate emission, like HD 50138, and another one exhibiting no silicate emission, but showing other small amplitude spectral features. A characteristic shape in the latter group consists of two wide bumps, one between 8 and $9.5\ \mu$m, the other between $10$ and $13\ \mu$m. We tentatively attribute this feature to either silica or to PAH emission \citep{vanBoekel2005}. In the group exhibiting silicate emission we find Herbig Be stars, such as HD 100546 and HD 259431, but also a supergiant B[e] star (HD 62623) and unclassified B[e] stars (HD 45677, HD 50138). The trend of the spectral shape with increasing baselines in HD 45677 is very similar to that of HD 50138, indicating similar inner disc structures. The other group, showing no or very weak silicate emission, also consists of Herbig Be (MWC 297), supergiant B[e] (GG Car, CPD-52 9243, CPD-57 2874), and unclassified B[e] stars (MWC 349 A, Hen 3-1191, MWC 300). Consequently, we can make no distinction on the evolutionary state of HD 50138 and other B[e] stars based only on the $8-13\ \mu$m spectral features. The material composition and structure of the circumstellar dust in these objects seem to be too diverse to be directly linked to the disc evolution. {Finally we note that HD 62623 and GG Car are confirmed binaries and CPD-52 9243 may be too \citep{Plets1995,Marchiano2012,Cidale2012}, and binarity possibly plays a key role in the B[e] phenomenon.}

	\section{Summary}
	We have studied the circumstellar disc of the unclassified B[e] star HD 50138 using mid-IR ($7.5-13\ \mu$m) interferometric spectra observed between 2007 and 2015  with the VLTI/MIDI instrument. The main results of our analysis are as follows:
	
	\begin{itemize}
		\item We determined the orientation of the circumstellar disc from the visibilities at $10.7\ \mu$m. The resulting inclination is $56.6^{\circ\, +2.3^\circ}_{\ \, -1.1^\circ}$, and the position angle is $63.4^{\circ\, +1.5^\circ}_{\ \,-3.4^\circ}$ (measured from north to east), in agreement with previous results.
		
		\item We fitted the interferometric data with a geometric disc model adopting ring geometry, and blackbody radiation with a power-law temperature profile. We obtained $2.7-3.7$~au for the disc inner radius, with an increasing trend with wavelength. The inner radius is larger than the dust sublimation radius, providing, to our knowledge the first detection of an inner {dusty} disc hole around a B[e] star. We note, however, that modeling of combined multi-wavelength data-sets are needed to confirm the detection.
		
		\item From the model we determine a range of $8.2-9.5$~au for the half-light radius as a function of wavelength. We found that the half-light radius at wavelengths corresponding to the continuum ($R_\mathrm{hl} = 8.2-8.6$~au) is smaller than at wavelengths corresponding to the silicate feature ($R_\mathrm{hl} = 9.2-9.5$~au). This means that the optically thick continuum is confined to a smaller area than the optically thin silicate emission.
		
		\item We analysed the MIDI spectra on different spatial scales. The correlated spectra show a clear feature between  $9.2$ and $11.3\ \mu$m. The amplitude of the feature gets smaller in the inner disc regions.
		
		\item The spectra  signs of dust evolution. We interpret the observed spectral shapes with two scenarios: (1) crystallization of silicate dust grains (forsterite, enstatite, silica), and (2) grain growth. We suggest that the spectral shapes in HD 50138 are more consistent with the presence of crystalline grains in the disc than with grain growth.
		
		\item 
		A novel result is that our interferometric data revealed a strong radial trend in the disc mineralogy: while the innermost region seems to include only forsterite grains, at intermediate radii both forsterite and enstatite are present. The outer disc may  predominantly contain amorphous silicate particles.
		
		\item Our observations implied no significant temporal variability of the mid-IR {disc morphology} or the silicate emission profile over the period 2007--2015. {However, we found that the average N band total flux was 30 per cent lower in 2007 than at the other epochs, which indicates real time variability.}
		
		\item We examined the MIDI data to find indications for binarity. We found no clear evidence for a companion which is farther than $\sim$50~mas, with mid-IR flux ratio larger than $\sim$$0.1$. 
		
		\item A comparison of the spectral shape of HD 50138 with a sample of various B[e] stars (supergiants, Herbig B[e], and unclassified B[e] stars) revealed that the evolutionary state of HD 50138 cannot be 
		determined from mid-IR spectroscopy.
		
	\end{itemize}
	
	\label{sec:con}
	
	{Further multi-wavelength observations are needed to better constrain the nature of this enigmatic object, especially the aspects of the variability, possible binarity, asymmetries in the disc structure and the origin of the crystalline silicate grains in the disc. The Multi AperTure mid-Infrared SpectroScopic Experiment \citep[MATISSE,][]{Lopez2014,Matter2016} at the VLTI will provide better $uv$-coverage compared to MIDI and it will measure closure phases in the N band for the first time. This will enable us to investigate possible binarity and disc asymmetries in great detail, with respect to earlier similar studies at near-IR wavelengths \citep{Kluska2016,Lazareff2017,Koutoulaki2018}. L and M band spectral coverage of MATISSE can also complement our view on the inner disc structure of HD 50138.}
	
	\section*{Acknowledgements}
	
	This project has received funding from the European Research Council (ERC) under the European Union's Horizon 2020 Research and Innovation programme under grant agreement No. 716155 (SACCRED). {This work is based on observations made with ESO telescopes at the Paranal Observatory. This work has made use of data from the European Space Agency
		(ESA) mission Gaia (https://www.cosmos.esa.int/gaia), processed by the Gaia Data Processing and Analysis Consortium (DPAC, https://www.
		cosmos.esa.int/web/gaia/dpac/consortium). Funding for the DPAC has been provided by national institutions, in particular the institutions participating in the Gaia Multilateral Agreement. The Combined Atlas of Sources with Spitzer IRS Spectra (CASSIS) is a product of the IRS instrument team, supported by NASA and JPL. KG acknowledges the J\'anos Bolyai Research Scholarship of the Hungarian Academy of Sciences. JV and KG acknowledge the Fizeau exchange visitors program which is funded by WP11 of OPTICON/H2020 (2017-2020, grant agreement 730890). We thank the anonymous referee for highly useful comments and suggestions.}
	
	
	
	
	\bibliographystyle{mnras}
	\bibliography{HD_50138_ref} 

	
	
	
	
	

	\bsp	
	\label{lastpage}
	
\end{document}